\documentclass[a4paper]{article}
\usepackage{bm, amsmath,amssymb,amsthm,amsfonts,graphics,graphicx,epsfig,rotating,caption,subcaption,url,natbib}
\newcommand{\para}{\bigskip\noindent}
\newcommand{\bbeta}{\bm{\beta}}
\newcommand{\expit}{\text{expit}}

\bmdefine\mmu{\mu}

\title{Better prediction by use of co-data: Adaptive group-regularized ridge regression}
\date{}
\author{Mark A. van de Wiel$^{1,2}$, Tonje G. Lien$^3$, Wina Verlaat$^{4}$,\\ Wessel N. van Wieringen$^{1,2}$, Saskia M. Wilting$^{4}$}

\begin{document}
\maketitle
\noindent
1. Department of Epidemiology and Biostatistics, VU University Medical Center, PO Box 7057, 1007 MB
Amsterdam, The Netherlands\\
2. Department of Mathematics, VU University, Amsterdam, The Netherlands\\
3. Department of Mathematics, University of Oslo, Oslo, Norway\\
4. Department of Pathology, VU University Medical Center, Amsterdam, The Netherlands\\



\para
\textbf{Supplementary Material is available from}: \url{www.few.vu.nl/~mavdwiel/grridge.html}\\

\begin{abstract}
\noindent
For many high-dimensional studies, additional information on the variables, like (genomic) annotation or external p-values, is available. In the context of binary and continuous prediction, we develop a method for adaptive group-regularized (logistic) ridge regression, which makes structural use of such `co-data'. Here, `groups' refer to a partition of the variables according to the co-data. We derive empirical Bayes estimates of group-specific penalties, which possess several nice properties: i) they are analytical; ii) they adapt to the informativeness of the co-data for the data at hand; iii) only one global penalty parameter requires tuning by cross-validation. In addition, the method allows use of \emph{multiple} types of co-data at little extra computational effort.

\para
We show that the group-specific penalties may lead to a larger distinction between `near-zero' and relatively large regression parameters, which facilitates post-hoc variable selection. The method, termed \texttt{GRridge}, is implemented in an easy-to-use R-package. It is demonstrated on two cancer genomics studies, which both concern the discrimination of precancerous cervical lesions from normal cervix tissues using methylation microarray data. For both examples, \texttt{GRridge} clearly improves the predictive performances of ordinary logistic ridge regression and the group lasso. In addition, we show that for the second study the relatively good predictive performance is maintained when selecting only 42 variables.
\end{abstract}

\section{Introduction}
Predicting binary or continuous response from high-dimensional data is a well-addressed problem nowadays. Many existing methods have been
adapted to cope with high-dimensional data, in particular by means of regularization and new ones, e.g. based on feature extraction, have been devised \cite[]{Hastie2008}. These methods have in common that the input is a response vector of length $n$ and a numerical $n \times p$ design matrix, where $n$ is the number of independent samples and $p>n$ is the number of variables. Then, the predictor is usually learned solely from this input, possibly in combination with or followed by variable selection.

\para
Co-data comprises of all information on the measured variables other than their numerical values for the given study. A few examples in the context of genomics are: a) Data or summaries like $p$-values from an external study with a related objective on the same set of variables (or highly overlapping); b) Database information that summarizes the (a priori) importance of genes for a class of diseases, e.g. the Cancer Gene census \cite[]{Futreal2004}; c) Genomic annotation, e.g. the chromosome on which a gene is located.
Co-data of type a), also referred to as `historical data', has been demonstrated to potentially benefit the analysis of a given clinical trial, in particular when sample size $n$ is small \cite[]{Neuenschwander2010}. For such low-dimensional data, assigning weight(s) to the co-data, e.g. by choice of the prior in a Bayesian setting, is a difficult issue, because it usually implies a subjective setting. In a high-dimensional setting like
ours, however, we show that one can use empirical Bayes principles to let the data decide how informative the co-data should be.

\para
The empirical Bayes approach sets our approach apart from other ones that use co-data to improve prediction or variable selection, like the group-lasso \cite[]{Meier2008}, a general multi-penalty approach \cite[]{Tai2007} or a weighted lasso approach  \cite[]{Bergersen2011}. In addition, unlike those methods, our approach is able to handle co-data of many different types: the external information on the variables can be binary, nominal, ordinal or continuous plus it can manage \emph{multiple} sources of co-data iteratively.

\para
We focus mostly on logistic ridge regression to present our approach, but also demonstrate the generality of the approach by an extension to random forest classification. We start out by recapitulating logistic ridge regression and the first two moments of the parameter estimates. These are then used to derive an empirical Bayes estimate for group-specific penalties. Next, we present a more stable iterative alternative, and also address iteration on multiple partitions of the variables. If the co-data is available as a continuous summary like a vector of $p$-values, we argue that one may use rank-based small groups of variables in combination with enforced monotony for the group-specific penalties.

\para
A consequence of the use of group-specific penalties is that it can facilitate \emph{a posteriori} variable selection. We show that effective group-regularization may result in a relatively heavy-tailed empirical distribution of the regression parameter estimates. This, as we illustrate by an example, may allow selection of a fairly sparse model with hardly any loss of predictive accuracy.

\para
The approach is demonstrated on two cancer genomics examples. Both examples concern discriminating precancerous cervical lesions from normal cervix tissues using methylation microarray data. For the first data set, we first demonstrate that our method can automatically account for different standard deviations across variables. Next, we show that the use of annotation of the methylation probes (which are the regression variables) for group-regularization improves the prediction for 86\% of the samples. The second example concerns a diagnostic setting using methylation profiles from self-collected cervico-vaginal lavages (self samples). The resulting samples are likely to be impure, which presents a challenge for discriminating the two classes. Here, we show that use of the $p$-values from the first study, which concerns more pure samples, as a basis for group-regularization in the second study, increases the area-under-the-ROC-curve from 67\% to 74\%. In addition, applying variable selection on the basis of the parameter estimates of the group-regularized approach rendered an equally accurate model with only 42 variables. Together with simulated data sets, the two examples were also used to compare our method with existing ones.

\para
We conclude with remarks on i) conceptual differences between our approach and related methods; ii) possible extensions of our method; and iii) the corresponding R-package \texttt{GRridge} and its computational efficiency.


\subsection{Logistic ridge regression}
It is well known that classical ridge regression corresponds to Bayesian ridge regression: the maximum a posteriori estimate for regression parameters $\bbeta = (\beta_1, \ldots, \beta_p)$ corresponds to the classical estimate $\hat{\bbeta}$ when using a central Gaussian prior for $\beta_k$ with a variance $\tau^2 \propto 1/\lambda,$ where $\lambda$ is the penalty parameter in the classical ridge setting.
We explore this fact to develop an empirical Bayes estimate of group-specific penalties.
We explain the procedure for logistic ridge regression; the changes needed for linear ridge regression are detailed in the Supplementary Material. The results of ordinary logistic ridge regression (hence ignoring the groups) at a given value of global penalty parameter $\lambda$ (e.g. obtained by cross-validation) are used as a starting point.


\para
We first recapitulate some results for logistic ridge regression. For independent responses $Y_i \in \{0,1\} , i=1, \ldots, n$, we have
$$Y_i \sim \text{Bernoulli}(\text{expit}(X_i\bbeta)),$$
where $X = (X^T_1, \ldots, X_n^T)^T$ is the $n \times p$ design matrix and $\text{expit}(X_i\bbeta) = \exp(X_i\bbeta)/(1+\exp(X_i\bbeta))$.
The estimate $\hat{\bbeta}$ maximizes the penalized log-likelihood:

\begin{equation}\label{loss}
\sum_{i=1}^n [Y_i \log(p_i) + (1-Y_i)\log(1-p_i)] - \lambda \sum_{k=1}^p\beta_k^2,
\end{equation}
where $p_i = \expit(X_i\bbeta)$.
Typically, the Newton-Raphson algorithm is used to maximize (\ref{loss}). Given current estimate $\tilde{\bbeta}$, define  $X_W = W^T X,$ $W = (\text{diag}(\tilde{p}_i(1-\tilde{p}_i)))^{1/2}$ and $\tilde{p}_i = \expit(X_i\tilde{\bbeta})$. Moreover, let $\mathbf{z} = (z_i)_{i=1}^{n}$ and $z_i = \text{logit}(\tilde{p}_i) + (Y_i-\tilde{p}_i)/(\tilde{p}_i(1-\tilde{p}_i)).$ Then, the Newton-Raphson update \cite[]{Cule2011} is:
\begin{equation}\label{ridgeest}
\hat{\bbeta} = (X_W^T X_W + 2\lambda I)^{-1} X_W^T \mathbf{z},
\end{equation}
and we assume (\ref{ridgeest}) has been applied until convergence.
Note that penalization causes bias, so, with $Y=(Y_1, \ldots, Y_n)$:  $E_Y(\hat{\beta}_k) \neq \beta_k$.
Both $E_Y(\hat{\beta}_k)$ and $V_Y(\hat{\beta}_k)$ can be approximated, as shown below. We will use these moments to derive an empirical Bayes estimate of the group-specific penalties.

\para
The first-order approximation $\mu_k$ of $E_Y(\hat{\bbeta})$ is \cite[]{leCessie1992, Cule2011}:
\begin{equation}\label{ridgeestapprox0}
\begin{split}
\mu_k &= [I - 2\lambda (X_W^T X_W + 2\lambda I)^{-1}\bbeta]_k = [(X_W^T X_W + 2\lambda I)^{-1}(X_W^T X_W + 2\lambda I - 2\lambda I)\bbeta]_k\\
&= [(X_W^T X_W + 2\lambda I)^{-1} X_W^TX_W \bbeta]_k =: \sum_{\ell=1}^p c_{k\ell}\beta_{\ell}
\end{split}
\end{equation}
where $[M]_k$ denotes the $k$th row (component) of any matrix (vector) $M$.
In addition, we have \cite[]{leCessie1992, Cule2011} for $\Sigma = \text{Cov}(\hat{\bbeta})$:
\begin{equation}\label{ridgevar}
\hat{\Sigma} \approx  (X_W^T X_W + 2\lambda I)^{-1} X_W^T X_W (X_W^T X_W + 2\lambda I)^{-1}.
\end{equation}
Calculation of both $\mu_k$ and $\hat{\Sigma}$ requires the inverse of the large $p\times p$ matrix $M_{\lambda} = (X_W^T X_W + 2\lambda I)^{-1}$. However, singular value decomposition (SVD) of $X_W^T = UDV^T$ reduces the calculation of $M_{\lambda}$ to inversion of an $n \times n$ matrix and matrix multiplication.

\subsection{Empirical Bayes estimation of group penalties}
Now, assume we have a partition of the variables into $G$ groups, $(\mathcal{G}_1, \ldots, \mathcal{G}_G)$, of sizes $(K_1, \ldots, K_G)$. Then, replace the penalty term in (\ref{loss}) by a generalized ridge penalty term \cite[]{Hoerl1970}:
\begin{equation}\label{loss2}
\sum_{i=1}^n [Y_i \log(p_i) + (1-Y_i)\log(1-p_i)] - \sum_{g=1}^G \lambda_{g} \sum_{k \in \mathcal{G}_g}\beta_{k}^2,
\end{equation}

\noindent
where $\lambda_g = \lambda'_g\lambda$ with global penalty $\lambda$ known and penalty multipliers $\lambda'_g$ to be estimated.
Let us assume an independent Gaussian (and hence ridge-type) prior:
\begin{equation}\label{prior}
\beta_k \sim N(0,\tau^2_{g(k)}),
\end{equation}
where $g(k)$ denotes the group that variable $k$ belongs to.
Then, for $E_{Y,\bbeta}(\hat{\beta}_k^2) = V_{Y,\bbeta}(\hat{\beta}_k)$, we have, using (\ref{ridgeestapprox0}) and the fact that $E_{\beta_i,\beta_j}\beta_i \beta_j = 0$,
\begin{equation}\label{var}
E_{Y,\bbeta}(\hat{\beta}_k^2) = E_{\bbeta}\left[V_Y(\hat{\beta}_k) + (E_Y(\hat{\beta}_k))^2\right]
= v_k + E_{\bbeta}[\mu_k^2] = v_k + \sum_{h=1}^G \sum_{\ell \in \mathcal{G}_h}c^2_{k\ell} \tau^2_h.
\end{equation}
Next, we obtain the $g$th estimation equation by 1) substituting $E_{Y,\bbeta}(\hat{\beta}_k^2)$ by its estimate, $\hat{\beta}_k^2$; 2) dividing both sides of (\ref{var}) by $v_k$; 3) subtracting 1 from both sides; and 4) aggregating over all $k \in \mathcal{G}_g$:
\begin{equation}\label{systemg}
\sum_{k \in \mathcal{G}_g}(\hat{\beta}_k^2/v_k - 1) = \sum_{k \in \mathcal{G}_g}v_k^{-1}\left[\sum_{h=1}^{G} \sum_{\ell \in \mathcal{G}_h}c^2_{k\ell} \tau^2_h\right] =  \sum_{k \in \mathcal{G}_g}\left[\sum_{h=1}^{G} \sum_{\ell \in \mathcal{G}_h}d^2_{k\ell} \tau^2_h\right] = : \sum_{h=1}^G \alpha_{gh}\tau^2_h,
\end{equation}
where $d_{k\ell} = c_{k\ell}/\sqrt{v_k}$. Let $B_g = \sum_{k \in \mathcal{G}_g}(\hat{\beta}_k^2/v_k - 1)$. Then, the empirical Bayes estimate for $\tau^2_1, \ldots, \tau^2_G$ is obtained by solving the system (linear in $\tau^2_h$):
\begin{equation}\label{system}
\left\{
  \begin{array}{ll}
    &B_1  = \sum_{h=1}^G \alpha_{1h}\tau^2_h \\
    &B_2  = \sum_{h=1}^G \alpha_{2h}\tau^2_h \\
    &.\\
    &.\\
    &B_G  = \sum_{h=1}^G \alpha_{Gh}\tau^2_h.
  \end{array}
\right.
\end{equation}

\para
Naive computation of the coefficients $\alpha_{gh}$ requires calculation of the possibly very large $p \times p$ matrix $D= (d_{k\ell})_{k,\ell=1}^p$. We experienced that this may consume considerable computing time and memory.
Fortunately, a much more efficient calculation is possible. To see this, first note from $d_{k\ell} = c_{k\ell}/\sqrt{v_k}$ and (\ref{ridgeestapprox0}) that $D$ is a product of an $p \times n$ matrix, $L=\text{diag}(1/\sqrt{v_k})(X_W^T X_W + 2\lambda I)^{-1} X_W^T$, and an $n \times p$ matrix $R = X_W$, where $L$ can be efficiently computed by SVD of $X^T_W$. Matrix decomposition of $L$ and $R$ according to the groups implies that $\alpha_{gh} = \sum_{k,\ell}(d_{k\ell}^{gh})^2$, where $d_{k\ell}^{gh}$ are the elements of $D^{gh} = L_g R_h$ with $L_g = (L_{k.})_{k \in \mathcal{G}_g}$ and $R_h = (R_{.\ell})_{\ell \in \mathcal{G}_h}$. $D^{gh}$ may still be a prohibitively large matrix, and hence we wish to avoid computing it. The following theorem provides an efficient solution for this when $p \gg n$, because it enables computation of $\alpha_{gh}$ by element-wise multiplication of $L^T_g L_g$ and $R_h R^T_h$, where both matrix products are of dimensions $n \times n$ only.

\para
{\bf Theorem} {\it Let $L$ and $R$ be $p_1 \times n$ and $n \times p_2$ matrices and $ D = LR$. Let $A \circ B$ be the Hadamard (element-wise) product of any equally-sized matrices $A$ and $B$. Denote the sum of elements of $A$ by $[A]_{\Sigma} = \sum_{k,\ell}a_{k\ell}$.  Then,}

\begin{equation}\label{ccomp}
\alpha = \sum_{k=1}^{p_1}\sum_{\ell=1}^{p_2}(d_{k\ell})^2 = [D \circ D]_{\Sigma} = [(L^T L) \circ  (R R^T)]_{\Sigma}.
\end{equation}
\noindent
{\bf Proof:} For any quadruple of matrices $A, B, C$ and $E$ of arbitrary dimensions $q \times r, q \times r, r \times s, r \times s$, respectively, we have \begin{equation*}\begin{split}[(A^T B) \circ (CE^T)]_{\Sigma} &= \sum_{k,\ell}\left(\sum_i (A^T)_{ki}b_{i\ell}\sum_j c_{kj}(E^T)_{j\ell}\right) = \sum_{k,\ell}\left(\sum_i a_{ik}b_{i\ell}\sum_j c_{kj}e_{\ell j}\right)\\ &= \sum_{i,j}\left(\sum_k a_{ik}c_{kj}\sum_{\ell} b_{i\ell}e_{\ell j}\right) = [(AC) \circ (BE)]_{\Sigma}.\end{split}\end{equation*}
Substituting $A=L, C=R, B=L$ and $E=R$ completes the proof. \hfill$\Box$

\para
System (\ref{system}) generally results in satisfactory solutions when $p$ is not extremely large (see also the Simulation section). For very large $p$, however, we experienced that it may lead to extreme (and even negative) values of the estimates. Such instabilities may be caused by strong multi-collinearity between variables (likely present in high-dimensional settings), also across groups, which affects the coefficients $\alpha_{gh}$ in (\ref{systemg}). Then, this may hamper disentangling the contributions of the various groups to each of the left-sides in (\ref{system}). Therefore, we provide an iterative alternative below, but we first discuss how to obtain the group penalties from
$\hat{\tau}^2_1, \ldots, \hat{\tau}^2_G$, the solutions of (\ref{system}).

\para
In order to allow re-estimation of $\bbeta$ the resulting group-specific variances, $\hat{\tau}^2_g$, are inverted to group-specific penalty multipliers $\lambda'_g$, which are calibrated towards the mean of their inverses equalling 1. This amounts to solving for constant $C$, with $K_g = |G_g|$:

\begin{equation}\label{empbayes5}
\lambda'_g = C/\hat{\tau}^2_g\ \ \text{and}\ \ \frac{1}{p}\sum_{g=1}^G K_g/\lambda'_g= 1.
\end{equation}
This calibration is useful to avoid (often time-consuming) re-cross-validation of $\lambda$.  It calibrates the mean of the inverse penalty multipliers towards the mean of those inverse multipliers in the original, initial ridge regression (with multipliers all equal to 1, implying a mean of 1). In fact, we observed for the examples below that after calibration re-cross-validation hardly changes the estimate of $\lambda$ and the predictive performance.
Finally, the group-specific penalty equals $\lambda_g = \lambda'_g \lambda.$

\subsection{Estimation for generalized logistic ridge regression}
After estimating the group-specific penalties we re-estimate $\bbeta$, which requires maximizing (\ref{loss2}). This is achieved by applying
ordinary logistic ridge regression, i.e. iteratively applying (\ref{ridgeest}), with penalty parameter $\lambda$ to a new weighted design matrix $X^{(2)}_W = X_W \Lambda^{-1/2}$, where $\Lambda$ is a diagonal matrix with $\Lambda_{kk} = \lambda'_{g(k)}$. To see this, write the group-specific penalty term corresponding to variable $k$ in group $g(k)$ as $$\lambda_{g(k)}\beta_k^2  = \lambda [(\lambda'_{g(k)})^{1/2}\beta_k]^2 =: \lambda (\beta'_k)^2.$$ Then, write the contribution of column $k$ in $X, [X]_k,$ to the penalized likelihood (\ref{loss2}) through $p_i = \expit(X_i\bbeta)$ as $[X]_k(\lambda'_{g(k)})^{-1/2}\beta'_k$, which determines $X^{(2)} = X \Lambda^{-1/2},$ and hence also $X^{(2)}_W =  W^T X^{(2)} = X_W \Lambda^{-1/2}.$ Finally, for the new estimate of $\beta_k$, we have:
\begin{equation}\label{newbeta}
\hat{\beta}^{(2)}_k = (\lambda'_{g(k)})^{-1/2}\hat{\beta}'_k.
\end{equation} Here, the upper index in $\hat{\beta}^{(2)}_k$ refers to the iteration, which will be introduced in the next section.  The variance should be scaled as well: $v^{(2)}_k = (\lambda'_{g(k)})^{-1} v'_k,$ with $v'_k = V(\hat{\beta}'_k),$ available from (\ref{ridgevar}).

\subsection{An iterative alternative}
Here, we provide an iterative alternative to (\ref{system}). The system (\ref{system}) does not make use of the fact that the initial estimates, $\hat{\bbeta}$, were implicitly (via the correspondence between the $\lambda$ and $\tau^2$) already obtained under a Gaussian prior with common variance $\tau^2$. In particular for high-dimensional data, this implicit prior has a large impact on $\hat{\bbeta}$. The proposed iterative alternative first estimates this common prior variance $\tau^2$. For that, we simply collect all variables in one group, which renders only one equation in (\ref{system}) with solution:
\begin{equation}\label{tau2}
\hat{\tau}^2 = \frac{\sum_{k=1}^p(\hat{\beta}_k^2/v_k - 1)}{\sum_{k,\ell=1}^p v_k^{-1} d^2_{k\ell}}.
\end{equation}

Then, we set out to estimate $\tau_g$ by first assuming $\tau^2_h = \hat{\tau}^2$ for all $h \neq g$, which is reasonable given the (implicit) common prior that was used to obtain the estimates. Now, splitting the right-side of the $g$th equation of (\ref{system}) into the contributions of group $g$ and all other groups and substituting $\tau^2_h = \hat{\tau}^2$ renders the estimate:
\begin{equation}\label{taug}
\hat{\tau}_g^2 = \frac{\sum_{k \in \mathcal{G}_g}(\hat{\beta}_k^2/v_k - 1) - \sum_{k \in \mathcal{G}_g}v_k^{-1}\sum_{h \neq g}\sum_{\ell \in \mathcal{G}_h}d^2_{k\ell} \hat{\tau}^2}{\sum_{k,\ell \in \mathcal{G}_g}v_k^{-1}d^2_{k\ell}}.
\end{equation}
In words, (\ref{taug}) can be considered as an estimate of $\tau_g^2$ that quantifies how much the observed sum of squared group $g$ parameters (scaled by their variances) deviates from the expected contributions to this summand of all variables $\ell$ not in group $\mathcal{G}_g$. The above solution is particularly attractive when iterating the estimation, because then the updated $\tau^2_g$ estimates adapt to the most recent generalized ridge estimates $\hat{\beta}'_k$ (\ref{newbeta}). As discussed above, these are also obtained under an implicit common prior (common $\lambda$), which allows us to iteratively use (\ref{tau2}) and (\ref{taug}). From $\hat{\beta}'_k$, the iterative re-scaling in (\ref{newbeta}) then computes $\hat{\beta}_k$, which is on the original scale of the covariates $X$. We experienced that this alternative solution is always very competitive with explicitly solving (\ref{system}), and sometimes superior, in particular when $p$ is (very) large. The Supplementary Material provides a simulation-based comparison between the two.

\para
Such iteration requires a stopping criterion. We simply monitor the cross-validated likelihood (CVL) and stop iterating when this decreases. The cross-validation is fast, because it only requires evaluation of the CVL for \emph{given} global penalty $\lambda$. Moreover, we use the efficient implementation by \cite{Meijer2013}. The resulting estimates are denoted by $\hat{\beta}^{(L)}_k$, where $L$ is the number of iterations before the CVL decreases.

\subsection{Iterating on a new partition}
More than one partition of the variables may be available, as illustrated in the second example. Suppose we have two partition with $G_1$ and $G_2$ groups, respectively. Then, the above method may simply be applied by cross-tabling the two partitions, rendering $G_1 G_2$ groups. However, this may render a very large number of groups and some of these groups may contain only few variables, which may deteriorate the empirical Bayes estimates.  Alternatively, one may simply iterate the group-specific regularization for the second partition after the first partition was considered. A disadvantage of that approach is that the results may (somewhat) depend on the ordering of the partitions. For the iterative re-penalization solution (\ref{taug}), we therefore opt to embed iteration on partitions into the re-penalization iteration. Hence, partitions are considered in alternating order. The CVL-based stopping criterion formulated above is applied to both partitions with respect to the previous fit; if CVL does not improve, that particular partition does not take part in the
outer re-penalization iteration anymore. If CVL does not improve for both partitions, the outer iteration is stopped as well. The group-regularization algorithm including this double iteration is depicted in Supplementary Figure 1.

\para
Note that the new penalty multipliers will adapt to both the data and the current penalties. This is important when the partitions are not independent. Let $\hat{\beta}^{(\ell,j)}_k$ be the estimate of $\beta_k$ for re-penalization iteration $\ell$ and partition $j = 1,2$. Then, the new estimate $\hat{\beta}^{(\ell,2)}_k$ is computed by applying (\ref{newbeta}) to $\hat{\beta}^{(\ell,1)}_k$, using grouping variable
$g_2(k)$ and $\hat{\beta}^{(\ell+1,1)}_k$ is computed by applying (\ref{newbeta}) to $\hat{\beta}^{(\ell,2)}_k$, using grouping variable
$g(k)$.  The final penalty multiplier for variable $k$ equals $\lambda'_{g(k)}\lambda''_{g_2(k)},$ where the latter term is the penalty multiplier based on the second partition. These notions trivially extend to more than two partitions. The final group-regularized estimates of $\beta_k$ are denoted by $\hat{\beta}^{\text{GR}}_k$. The iterative group-regularization is illustrated in the second example.

\subsection{Ranking-based groups}
Often, the co-data consist of external data on the same variables (e.g. genes) for an analogous, but somewhat different setting. Our second data example illustrates such a case. Then, the two data sets can not simply be pooled. However, summaries like $p$-values or regression coefficients based on the external data may be used to define a partition of the variables into small groups which is then used as input for the group-regularized ridge on the primary data set. We enforce monotony on the penalties of those groups to avoid over-fitting, as detailed below.

\para
First, rank the variables according to the summary, e.g. $p$-values. Then, create groups of size $s$, where group $g$ contains the variables with ranks $s(g-1) + 1, \ldots, sg$. Apply (\ref{taug}) to obtain initial estimates $(\hat{\tau}^{\text{init}}_{g})^2$ for these small groups. Due to the size of the groups these estimates may be instable and not in line with the ranking based on the external data. Therefore, we force the estimates to be monotone by applying weighted isotonic regression \cite[]{Robertson82} of $(\hat{\tau}^{\text{init}}_{g})^2$ on the index (and hence group rank) $g$, rendering regression function $\hat{f}()$. The weights account for possibly different group sizes. Then, the new estimates are set to $\hat{\tau}^2_{g} = \hat{f}(g)$, which are substituted into (\ref{empbayes5}) to obtain group-specific penalty multipliers $\lambda'_g$. Enforcing monotony highly stabilizes the estimates and interpretation of the results. In fact, even $s=1$ might be used, but, because the stabilizing effect of the isotonic regression is potentially less strong for the extreme ranks, this could lead to over-fitting. The latter is mitigated by using small, non-singular groups. The stabilizing effect is illustrated for the second data example in Supplementary Figure 2.

\para
The software also allows for non-uniformly-sized groups, where one specifies a minimum group size, say $s=10$, for variables corresponding to the most extreme values of the summary, and a maximum number of groups; the group size then gradually increases for variables with less extreme values of the summary. This enables the use of fewer groups (and hence faster computations) while still maintaining a good `resolution' for the extreme values of the summary.


\section{Generalizing the concept I: post-hoc variable selection}
A nice side effect of group-specific regularization is that it may simplify post-hoc variable selection,
because the empirical distribution of estimated coefficients is typically more heavy-tailed than the one from ordinary ridge regression. Hence, there is a clearer separation between $\hat{\beta}_k$'s close to zero from those further away from zero.  This is illustrated in Supplementary Figure 4 for the second data example. Also, it is known that ordinary ridge regression tends to spread mass of the parameter estimates over correlated variables.
Group-specific regularization can prioritize such variables, in particular when the groups are small and the range of group-specific penalties is large. A posteriori selection could be based on an information criterion or a mixture model for the $\hat{\beta}_k$'s. However, since we are in a prediction setting, we suggest to select directly on the basis of predictive performance by using CVL. For the purpose of prediction, variable selection is mainly desirable for potentially developing a measurement devise (e.g. based on qPCR) with much fewer variables than the original one. Hence, we allow the user to set a maximum of variables to be selected, e.g. $p_\text{max} = 100$.

 \para
A simple proposal for CVL-based selection is: sort the variables with respect to $|\hat{\beta}^{\text{GR}}_k|$; select $s, 0 \leq s \leq s_{\text{max}}$ top-ranking variables; re-fit the model using only those variables, but with the same fixed $\lambda$ and $\lambda_g$'s as for the full model; compute $\text{CVL}_s$ on this model; find $\text{CVL}_\text{max} = \max_s \text{CVL}_s$; select $s_{\text{sel}} = \min\{s : \text{CVL}_s \geq \text{CVL}_\text{max} - q_{\text{marg}}|\text{CVL}_\text{max}|\}$, with e.g. relative margin $q_{\text{marg}} = 1\% = 0.01$. The margin favors more sparse models: the minimization finds the model with the fewest variables such that its CVL is within a, say, 1\% margin of the best. Supplementary Figure 1 depicts the entire group-regularization algorithm including variable selection, whereas Supplementary Figure 5 shows the CVL profile as a function of $s$ for the second data example.

\section{Generalizing the concept II: random forest}
The concept of adaptive group-regularization (or, analogous, group-weighting) can be generalized to other classifiers, also to some of very different nature than logistic ridge regression. The Supplementary Material describes the extension to the random forest classifier in detail; below we provide a summary.

\para
A standard random forest classifier uses only $m = \mathcal{O}(\sqrt p)$ variables (nodes) per node split. Typically, these variables are sampled uniformly from the entire set. Now, the idea is to weigh groups by increasing or decreasing the sampling probability according to the overall importance of variables in a group. Once a set of top-ranking variables across a forest is defined by a formal selection procedure \cite[]{Doksum2008} or by simply using the top $k\%$ (for, say, $k=5$), the observed number of top-ranking variables per group is modeled by a multinomial distribution per tree. Then, the variability of the multinomial proportions across trees is modeled by a Dirichlet distribution the parameters of which are estimated by use of empirical Bayes. This Dirichlet distribution is then used for weighted sampling of variables in the trees in a new random forest. The process of random forest classification, variable ranking, selection, estimation and weighted sampling is repeated, until the out-of-bag error does not or hardly decrease anymore.

\section{Simulation results: summary}
We performed simulations to compare the performances of the systems-based solution (\ref{system}) and the iterative solution
(\ref{taug}). In addition, \texttt{GRridge} is compared with i) ordinary logistic ridge and ii) the group lasso \cite[]{Meier2008}. We also compared with the adaptive logistic ridge, which is the ridge version of the adaptive logistic lasso \cite[]{Zou2006}, simply amounting to using variable-specific penalty multipliers that are inverse proportional with the initial squared ridge parameter estimates, $\hat{\beta}_k^2$. However, since we found that the predictive performance of the adaptive logistic ridge was generally inferior to that of the ordinary logistic ridge, we do not present those results in detail.

\para
We study a number of scenarios where we vary the number of groups $G$, the size of the groups $p_g$, the correlation strength in
$X$, the differential signal between the two classes of samples across groups, and the sparsity (i.e. proportion of groups without predictive signal). Performance was evaluated by computing AUC and mean Brier residuals on a large test data set ($n_{\text{test}}=1000$), which was generated under the same settings as the training set ($n=100$). These are reported in extensive tables, supplied in the Supplementary Material; here, we summarize the results.

\para
First, we observe that the systems-based solution and the iterative solution are very competitive for $p = 2000, 5000$ ($p = G*p_g$), while
the latter is superior for large $p, p = 12500.$ In particular, the iterative solution is indeed more stable across repeated simulations for very large $p$. The non-iterative, systems-based solution relies strongly on the parameter estimates of the initial logistic ridge regression, the bias of which may be very strong when $p$ is very large. The iterative solution, however, typically finds less extreme group penalties in the first iteration, then  re-estimates the regression parameters, allowing those to adapt to the new penalties.

\para
Generally, both \texttt{GRridge} versions performed at least as good as ordinary logistic ridge. As expected, we clearly observe that the gap between the performances increases with more skewed effects across groups and with increased sparsity. In addition, group lasso outperforms ordinary logistic ridge in group-sparse settings, while the reverse often holds for the non-sparse settings. \texttt{GRridge} generally outperforms the predictive accuracy of the group lasso, in some cases with fairly large margins, e.g. with AUCs that are 0.10 to 0.15 larger on the absolute scale. The group lasso becomes more competitive for high group-sparsity, in particular for $p$ large. Yet, it seems that \texttt{GRridge} adapts well to sparsity and maintains its relative good performance. Note that the weaker predictive performance of the group lasso may, for some applications, be counterbalanced by its group-selection property.

\section{Examples: diagnostic classification using methylation data}
DNA consists of the four nucleotides A, C, G and T. Methylation refers to the addition of a methyl-group to a C preceding a G (CpG), which can influence expression of the encoded gene. As such, methylation has a so-called epigenetic effect on the functionality of the cell, and consequently on the entire organism. It is believed to be an important molecular process in the development of cancer \cite[]{Laird2003}. In addition, DNA is a well-characterized and relatively stable molecule, compared to mRNA (gene expression) and many proteins. Therefore, the use of DNA methylation for diagnostic purposes is currently heavily investigated. A popular platform for measuring methylation is the Illumina\texttrademark 450K bead chip. This platform measures 450.000 probes per individual, where each probe corresponds uniquely to a CpG location on the genome. Each probe renders a so-called beta-value, which is the estimated proportion of methylated DNA molecules for that particular genomic location in a given tissue. Like for any microarray study, the data is preprocessed using several steps; see the Supplementary Material.

\para
We have data sets from two similar studies on cervical cancer at our disposal. The carcinogenesis of cervical cancer is relatively well-characterized. The transformation process of normal epithelium to invasive cancer takes many years, and includes distinct stages of precursor lesions (CIN; cervical intraepithelial neoplasia). Whereas low-grade precursor lesions (CIN1/2) are known to regress back to normal, high-grade precursor lesions (CIN2/3) have a relatively high risk for progression to cancer and are usually surgically removed. Therefore, accurate detection of high-grade CIN is very important. The two studies both measure methylation for normal cervical tissue and CIN3 tissue for several independent individuals, but differ in one important aspect. The first study measures methylation on CIN3 tissue biopsies, whereas the second study considered self-collected cervico-vaginal lavages of women with underlying CIN3 lesions \cite[]{Gok2010}.
The relatively good quality of the samples in the first study may render important information about relevant methylation markers. The quality and purity of the tissues in the second study is probably inferior. This study, however, better resembles a more realistic diagnostic setting, in particular because many countries have implemented screening programs for cervical cancer. Our first example uses the data of the first study only, but compliments this with another source of information: annotation of the probes.  This creates a partition of the probes into groups, which is used in the group-regularized ridge regression. The second example shows how, in addition to the annotation, the results of the first study can be used in our algorithm to improve diagnostic classification for the second study. We present the results from the iterative version of our method.

\subsection{To standardize or not? - An automatic solution}
A practical issue when applying penalized regression is the need or `no need' for standardization of the covariates. There is no consensus on this issue \cite[]{Binder2014}, because on one hand standardization has the beneficial effect of rendering a common penalty more appropriate, while on the other hand it may remove some of the (differential) signal and may lead to instabilities for  variables for which the sample variances are small. Standardization is equivalent to introducing a penalty multiplier that is  proportional to the variance in the unstandardized setting \cite[]{Binder2014}. A potential of our method is that it can let the data decide how the variances of the variables should impact the penalties. Below, we explore this potential for the first study.

\para
The first study \cite[]{Farkas2013} contains methylation profiles of 20 and 17 unrelated normal cervical and CIN3 tissues, respectively. To enable inclusion of these data in our complementary R-package \texttt{GRridge}, thereby allowing reproduction of the results, the computations for this and the next example were performed on a random selection of 40,000 probes. We verified, however, that all results are very similar on the entire data set, which is not surprising given the smooth nature of ridge regression and the correlations between variables.

\para
The probes are grouped in 8 groups of 5,000 each, in increasing order of the sample variances. Note that we verified whether a different grouping (e.g. 16 groups of 2,500 or 40 groups of 1,000) would affect the results. This is not the case. In line with the argumentation above (larger penalties for probes with large variances) we imposed monotony on the 8 penalty multipliers. We observed that this constraint is not very essential here, because the estimated penalty multipliers are also nearly monotonously increasing when the constraint was not imposed. \texttt{GRridge} estimated the following
penalty multipliers: $(2.75*10^{-2}, 6.59*10^{-2}, 6.92*10^{-2}, 8.03*10^{-2}, 1.21*10^{-1}, 3.00*10^{-1}, 7.36*10^8, 2.75*10^9)$. So, it effectively
completely removes the impact of the probes with large variances (last two groups), allowing smaller penalties for the remaining 6 groups. Interestingly, \texttt{GRridge} with variance-based groups outperforms \emph{both} ordinary standardized and unstandardized logistic ridge, in terms of ROC-curves (see Supplementary Material), AUC (0.91, 0.86, 0.76, respectively) and mean Brier residuals (defined as $1/n \sum_{i=1}^n (Y_i - p_i)^2$; 0.14, 0.16, 0.21, respectively).

\subsection{Improved classification by use of probe annotation}
Our hypothesis here is that the use of a priori known annotation-based partitions of the probes may improve the classification results. This second partition, next to the variance-based one above, is based on the probe's location in or nearby a so-called CpG-island. A CpG-island is a genomic region which is relatively rich in CG base pairs, and methylation is known to be more prevalent there than elsewhere. We used the following groups (in order of decreasing distance to CpG-islands): ``CpG-island (CpG)'', ``North Shore (NSe)'', ``South Shore (SSe)'', ``North Shelf (NSf)'', ``South Shelf (SSf)'', and ``Distant (D)''. If probes in CpG (or any other group of probes) are on average more important for the classification, the group-regularized ridge automatically detects this and applies a smaller penalty to all probes in this group. This may improve classification when the a priori partition was indeed informative. Note that the partition used is based on a well-accepted criterion to characterize genomic locations in methylation studies.


\para
The group-regularized ridge used 6 iterations for re-penalizing the 6 annotation-based groups and 7 iterations for the 8 variance-based ones, which increased the CVL by 40\% from -20.18 to -12.03. The final penalty multipliers for the annotation-based groups ($\propto$ inverse weights) are: $\lambda'_{\text{CpG}} = 0.015$, $\lambda'_{\text{NSe}} = 278$, $\lambda'_{\text{SSe}} = 0.12$, $\lambda'_{\text{NSf}} = 2,986$, $\lambda'_{\text{SSf}} = 2,987$ and $\lambda'_{\text{D}} = 685$. The group-specific penalties clearly affect the regression parameter estimates $\hat{\beta}_k^{\text{GR}}$, because larger values of $\lambda'_{g}$ result in smaller values of $|\hat{\beta}_k^{\text{GR}}|$. Hence, these values imply that \texttt{GRridge} effectively only uses the CpG and SSe probes for the predictions. The results confirm the importance of probes on CpG islands. The variance-based penalty multipliers are $(1.93*10^{-1}, 2.41*10^{-1}, 2.41*10^{-1}, 2.41*10^{-1}, 2.92*10^{-1}, 6.74*10^{-1}, 3.29*10^{4}, 1.11*10^{5})$. These are largely in line with the results above, although somewhat compressed, because they adapted to the annotation-based multipliers. Please note that one should be careful with interpreting the exact values of the group penalties. As indicated above, these may depend on the presence of another partition due to overlap between groups. In addition, in the Supplementary Material we show that for the annotation-based groups above the penalties vary somewhat with respect to sizes of the groups. The order of the group-penalties seems to be fairly stable, however, so we recommend to interpret the group-penalties in terms of their ranking.

\para
To assess whether the group-regularized ridge improves classification with respect to ordinary ridge, we computed ROC curves obtained by 10-fold cross-validation. Here, we compare with ridge regression on standardized covariates, because the latter was superior to the unstandardized version, as demonstrated above. In addition, we compare with the group lasso \cite[]{Meier2008}, as implemented in the R package \texttt{grpreg}, using the same annotation-based groups as for \texttt{GRridge}. The last competitor is the adaptive ridge, as discussed above. Also these two methods are applied to the standardized covariates (which were verified to be superior to their unstandardized counterparts). Group lasso selected only the SSe group, but, surprisingly, not the CpG group.

\begin{figure}[h]
\begin{subfigure}[h]{0.5\textwidth}	
\includegraphics[width=\textwidth]{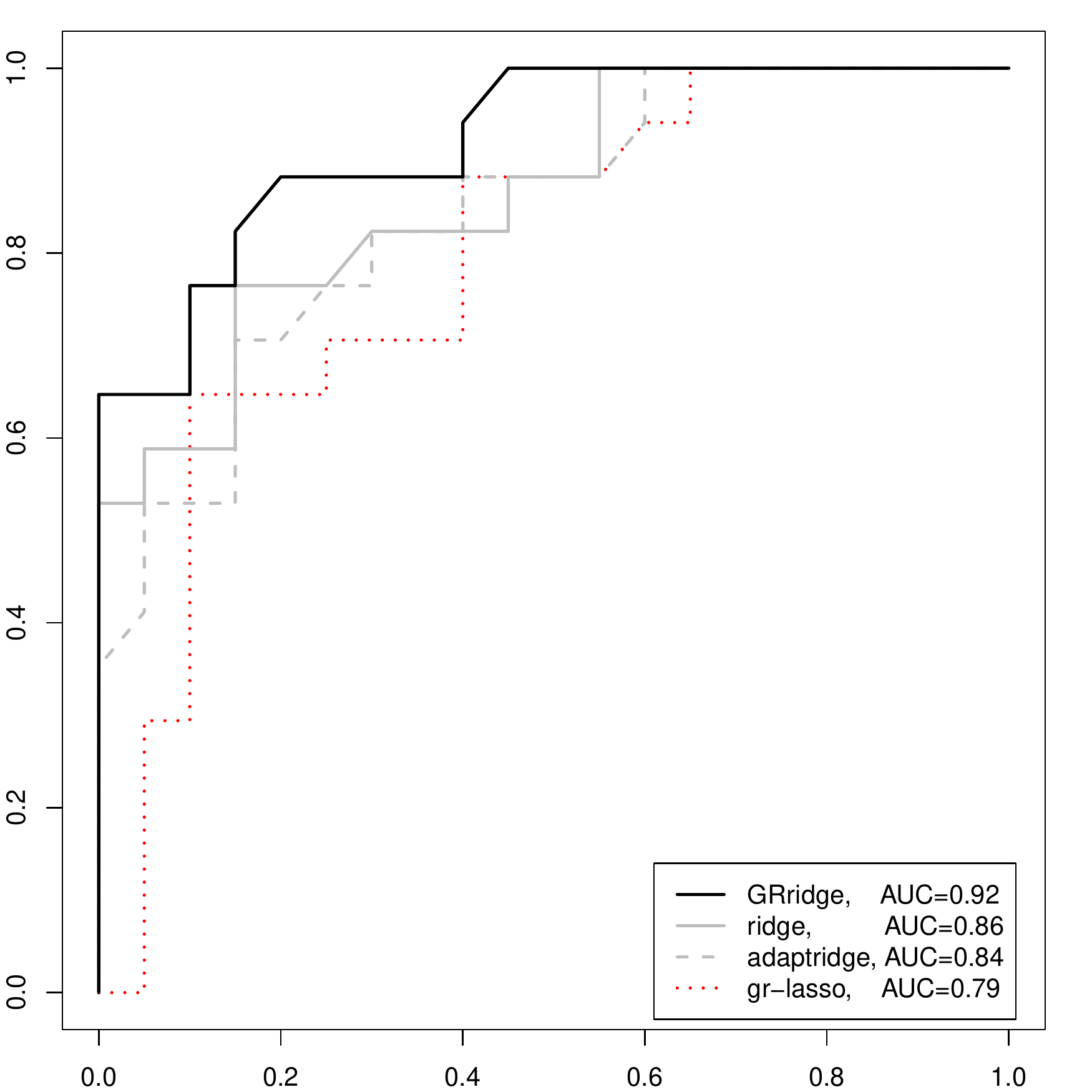}
\caption{}
\end{subfigure}
\begin{subfigure}[h]{0.5\textwidth}
\includegraphics[width=\textwidth]{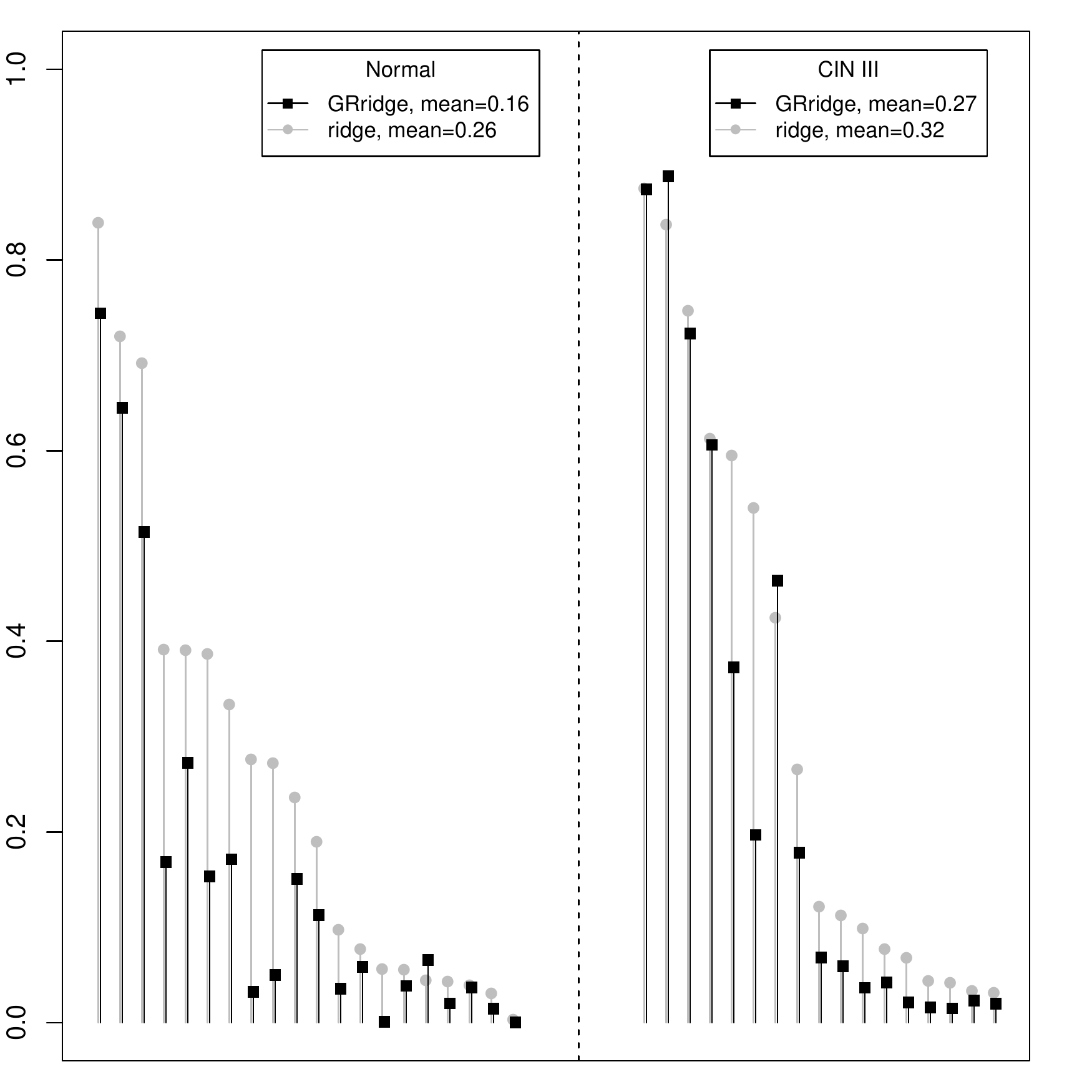}
\caption{}
\end{subfigure}
\caption{(a): ROC curves for first example, Group-regularized ridge (\texttt{GRridge}), ordinary \texttt{ridge}, group lasso \texttt{gr-lasso} and adaptive ridge \texttt{adaptridge}. X-axis: False Positive Rate, y-axis: True Positive Rate. (b): Absolute residuals $|Y_i - p_i|$ for both classes for \texttt{GRridge} and \texttt{ridge}, in decreasing order of the \texttt{ridge} residuals}
		\label{rocsfarkas}
\end{figure}

\para
The resulting ROC curves depict the False Positive Rate (FPR) versus the True Positive Rate (FPR) for a dynamic cut-off for the predicted probability on CIN3. Figure \ref{rocsfarkas}(a) shows the results.  We clearly observe superior performance of \texttt{GRridge}, with  AUC = 0.92 (and 0.86, 0.84, 0.79 for ordinary ridge, adaptive ridge and group lasso, respectively).
With respect to ordinary ridge, predictions improved for 33 out of 37 observations, as displayed in Figure \ref{rocsfarkas}(b). Note that adding the annotation-based groups to the variance-based groups improved the AUC only slightly, from 0.91 to 0.92, probably because AUC is a rank-based criterion. In fact, the predictions did improve for 32 out of 37 observations, leading to a relatively larger improvement (decrease) for the mean Brier residual, from 0.143 to 0.116.




\subsection{Improved diagnostic classification by use of external data}
The second study contains methylation profiles of self-collected cervico-vaginal lavages (or self-samples) corresponding to 15 women with an unaffected (normal) cervix and 29 women with CIN3 lesions, all unrelated. Here, it is important to note that the samples of the affected cervices may be contaminated with normal cells and cells of other origins (mostly vaginal cells and lymphocytes), due to imprecise sampling. Hence, the differential signal may be diluted. We aim to use the results of the first study for the group penalties in the second study.

\para
In principle, we could use the results of the group-regularized ridge regression fitted on the first study, as presented in the previous section. However, the effect of the (possible) contamination may vary considerably across probes. For example, the differential signal of probes with hypo-methylation (affected $<$ normal) in the first study is diluted more than that of hyper-methylated probes. This can be illustrated in a simple deterministic setting. In case of hypo-methylation, consider a true ratio affected/normal = 0.4/0.8 = 1/2. Assume a contamination of 50\%, then the measured ratio will be (0.4/2 + 0.8/2)/0.8 = 3/4, hence the ratio is 50\% too large. Using the same numbers for hyper-methylation renders a measured ratio that is only 33\% too small. In addition, it is well-known that ridge regression distributes differential signal over parameters corresponding to correlated probes. Hence, the magnitude of a particular coefficient also depends on other probes. Since the dilution in Study 2 affects probes differently, the applicability of Study 1 ridge regression results for analyzing Study 2 may be limited.

\para
Therefore, we propose to use group penalties $\lambda_g$ that are simply based on $t$-test $p$-values as obtained by applying limma \cite[]{Smyth2004} on Study 1. These $p$-values are then used to define a ranking-based partition with 100 groups of probes of minimal size $s=10$ (size gradually increasing with the $p$-value) as described above. To stabilize the estimates of $\lambda_g$ weights $\hat{\tau}^2_g \propto 1/\lambda_g$ for Study 2 are forced to be monotonously decreasing with increasing Study 1 $p$-values as described above. The function \texttt{pava} of the R library \texttt{Iso} is used for this purpose, which is illustrated in Supplementary Figure 2. In this setting, it is reasonable to precede our method by a mild \emph{prior} filtering: only include those probes with $\text{FDR} \leq 0.5$ and a mean absolute difference larger than 0.1 (on log-scale) \emph{in Study 1}. Then, our method applies group-specific regularization to the 9491 probes surviving these thresholds.

\para
Given the earlier argument about a stronger dilution effect on hypo- than on hyper-methylated probes (as detected in Study 1), we also considered a second sign-based partition that distinguishes those two groups of probes. Finally, we added the variance-based and annotation-based partitions introduced in the first example. This illustrates the ability of our method to operate on multiple partitions. For this example, the adaptive group-regularized ridge used 3 re-penalization iterations. The CVL increased from -28.91 to -27.54, hence a 5\% improvement. The sign-based and variance-based partitions had no effect on the results (hence rendering group-specific penalties equal to 1) on top of the $p$-value-based partition, illustrating the adaptive nature of the algorithm.  The partition based on external $p$-values produced 100 group-specific penalties ranging from $1.3*10^{-3}$ to $13.3$ for $g = 1, \ldots, 100$, so indeed a large range (see also Supplementary Figure 3), illustrating the relevance of this partition.  The annotation-based partition
rendered $\lambda'_{\text{CpG}} = 0.17$, and much larger penalty multipliers for the other 5 classes. So, also for this data set the probes on the CpG-islands correspond to smaller penalties, which is biologically plausible.


\para
We compared \texttt{GRridge} with: i) ordinary ridge; ii) adaptive ridge; iii) group lasso with the same 100 $p$-value based groups (all three also applied to the filtered probe set); and iv) ordinary ridge on the entire, non-filtered probe set.
For this data set, the group lasso did not select any group.
Hence, no variable was selected either, rendering inferior prediction results. Probably, the weak differential signal per variable in this challenging data set caused the absence of selections for the group lasso. The ROC curves for the other three methods were obtained by applying leave-one-out cross-validation (LOOCV). Figure \ref{rocs}(a) shows the results: \texttt{GRridge} has a markedly higher AUC (0.74) than the ones corresponding to i) 0.67, ii) 0.67 and iv) 0.63.

\para
We also checked whether the order in which the four partitions are used within each re-penalization iteration matters for the results. The final CVLs for all 24 possible orderings show very little variation: the range is $[-24.58, -24.54]$. Hence, we conclude that the sensitivity of the performance with respect to the ordering is very small for this data set.

%

\subsection{Variable selection}
\begin{figure}[h]
\begin{subfigure}[h]{0.5\textwidth}	
\includegraphics[width=\textwidth]{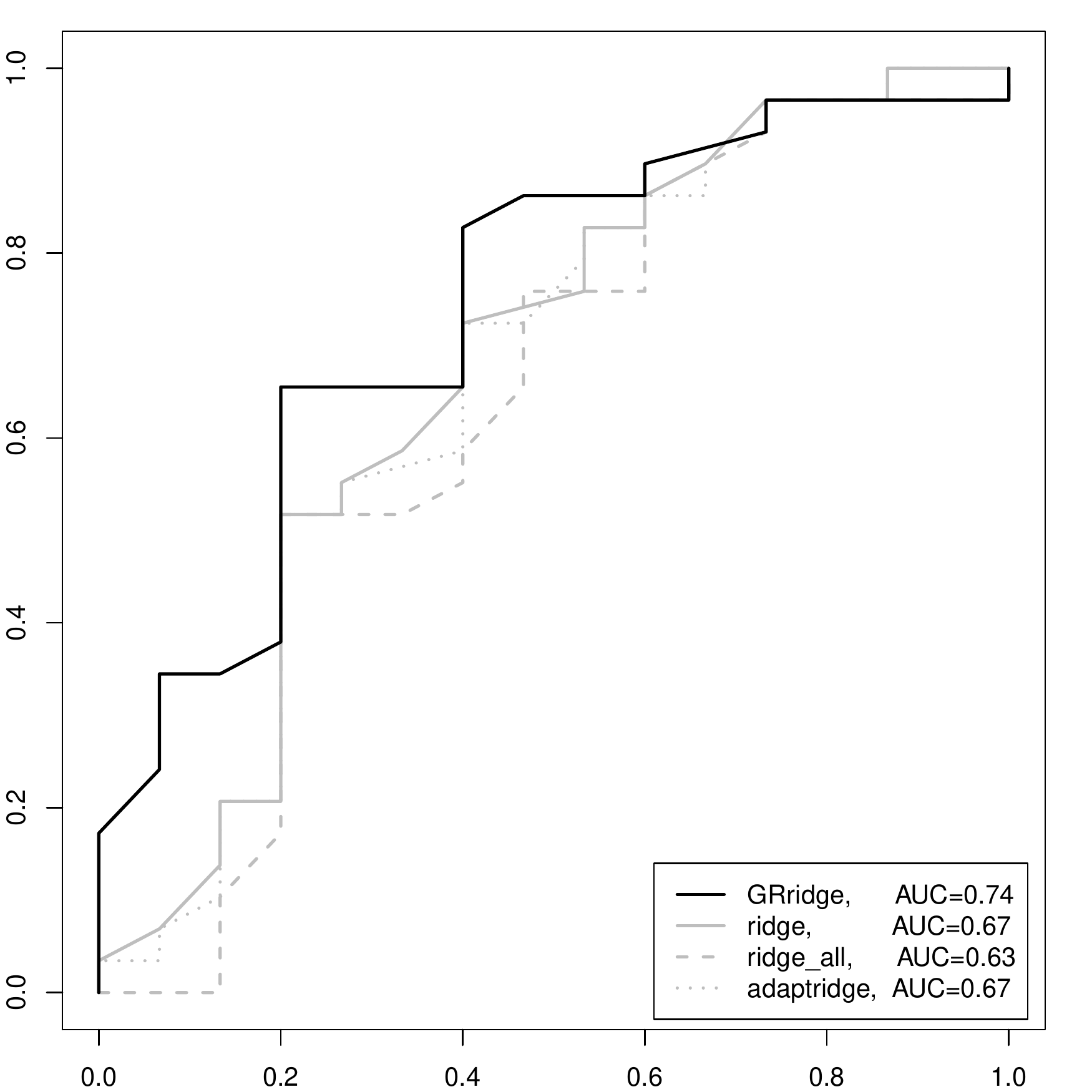}
\caption{}
\end{subfigure}
\begin{subfigure}[h]{0.5\textwidth}
\includegraphics[width=\textwidth]{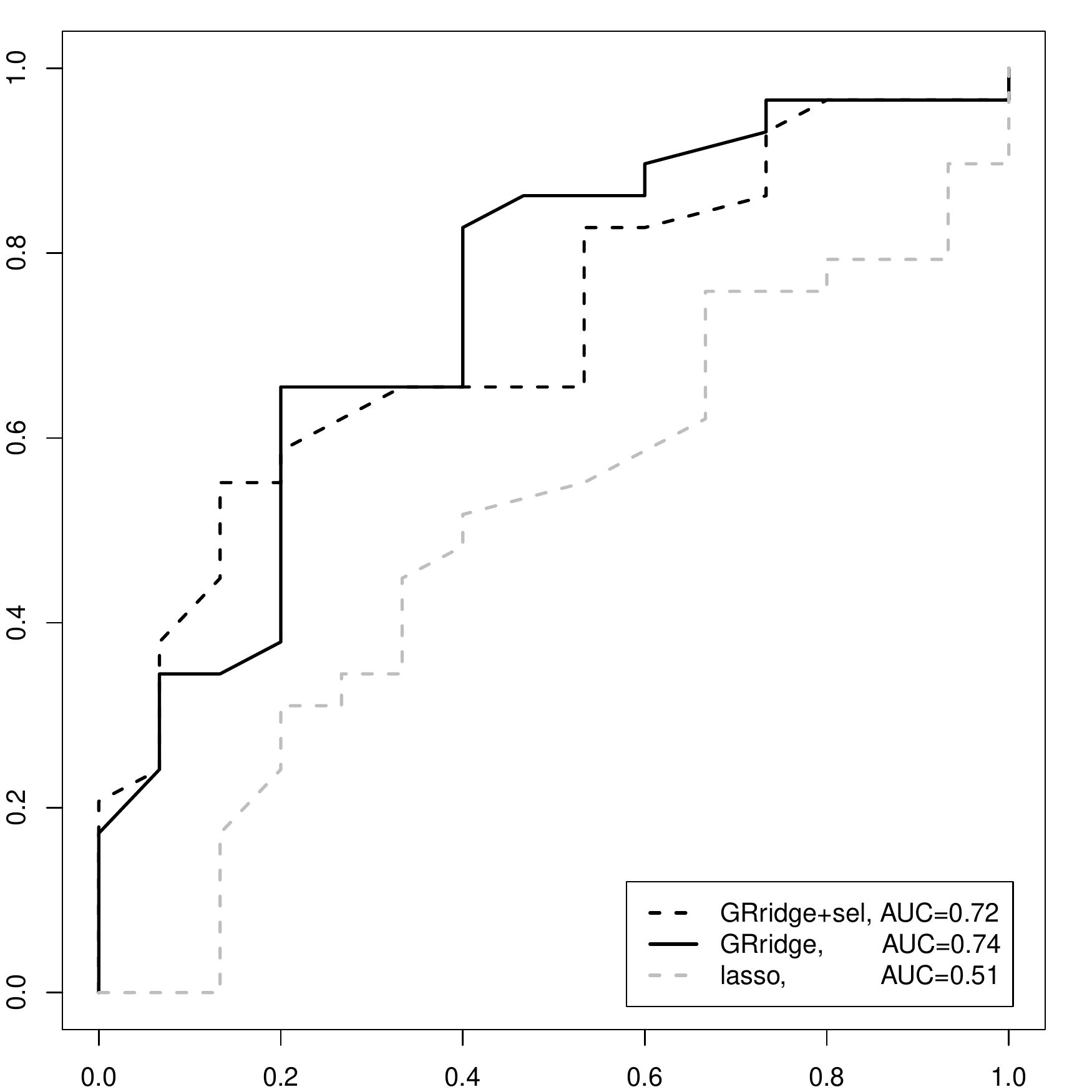}
\caption{}
\end{subfigure}
\caption{ROC curves for second example. Sub-figure (a): group-regularized ridge (\texttt{GRridge}), \texttt{ridge} and ridge on all variables (\texttt{ridge\_all}); (b): Group-regularized ridge plus variable selection (\texttt{GRridge+sel}), \texttt{GRridge},  and \texttt{lasso}. X-axis: False Positive Rate, y-axis: True Positive Rate.}
		\label{rocs}
\end{figure}

Supplementary Figure 4 shows that the most extreme coefficients of the group-regularized ridge regression are relatively much larger than those of ordinary logistic ridge regression. In fact, for the former, the 1\% most extreme coefficients account for 61\% of the total sum of absolute values of the coefficients, whereas for the latter this drops to only 3\%.
We applied the proposed \emph{a posteriori} variable selection to $\hat{\beta}^{\text{GR}}_k$ which rendered a model with 42 selected variables, termed \texttt{GRridge+sel}. Figure \ref{rocs}(b) depicts the ROC-curves and AUCs for \texttt{GRridge+sel}, \texttt{GRridge} and \texttt{lasso}, as obtained by LOOCV. First, note that the much more parsimonious \texttt{GRridge+sel} model predicts nearly as well as the full \texttt{GRridge} model in this case (AUC = 0.72 vs AUC = 0.74). Second, to illustrate the beneficial effect of group-specific regularization in this variable selection context, we also compare \texttt{GRridge+sel} with the \texttt{lasso} \cite[R package \texttt{penalized}]{Goeman2010b} on the same filtered data set. The \texttt{lasso} renders a somewhat more parsimonious model with 17 variables, but performs much worse in terms of prediction: Figure \ref{rocs}(b) depicts the ROC-curves and AUCs. Of course, the \texttt{lasso} could possibly be improved by adapting group-regularized principles as well (see Discussion).

\section{Discussion}
Our method is weakly adaptive in the sense that the penalties adapt in a group-specific sense only.  This is an important conceptual difference with strongly adaptive methods such as adaptive lasso \cite[]{Zou2006} and enriched random forests \cite[]{Amaratunga2008}, which aim to learn variable-specific penalties from the same data as the data used for classification. Such methods strongly rely on sparsity. While this may be a fairly natural assumption for some applications, we believe it to be less realistic for complex genomic traits like cancer. In fact, we observed that for both applications the adaptive lasso did not outperform the ordinary lasso, and hence performed worse than the adaptive group-regularized ridge regression.

\para
The adaptive group-\emph{regularized} ridge shares the philosophy of accounting for group structure with the group lasso \cite[]{Meier2008}. The latter, however, \emph{selects} entire groups using a lasso penalty on the group-wise sum of coefficients and then spreads the coefficients within a group using a ridge penalty within a group. The group lasso is particularly attractive for selecting relatively small, interpretable groups of variables, e.g. gene pathways. However, it is less useful and suitable when the groups tend to be large (and not necessarily homogenous) as in the first example, or when the groups have no clear biological interpretation, as for the ranking-based small groups in the second example. In addition, for both simulated and real data sets we show that the predictive performance of \texttt{GRridge} is often superior to that of the group lasso.
Group-specific regularization was also discussed by \cite{Tai2007} in the context of nearest shrunken centroids and partial least squares classifiers. Their results support our claim that such regularization can improve classification performance. Their approach, however, requires cross-validation on \emph{all} group-penalties or, when this is too computationally demanding, \emph{a priori} fixing of weights (inverse penalties). Also, unlike \texttt{GRridge}, their method does not make use of multiple partitions of the variables, which are often available in practice.

\para
As discussed, group-regularization helps to better discriminate small and large coefficients, and the model after variable selection may be fairly parsimonious.
Yet, extension of our method to sparse methods like lasso may be desirable in some cases. These methods usually render only few non-zero coefficients, which may lead to unstable group penalties. This may be mitigated by re-sampling or by using a power transformation of ridge-based penalties, as suggested by \cite{Bergersen2011} in another setting. Alternatively, one may consider a Bayesian set-up with a selection prior, for example a Laplace prior \cite[]{ParkCasella2008} or a horseshoe prior \cite[]{CarvalhoCarlos2009}. The hyper-parameters of such priors would be estimated \emph{per group of variables}, e.g. by empirical Bayes. Then, the entire posterior of each $\beta_k$, rather than just the point estimate, impacts the penalty (represented by the group-wise prior) of the group that variable $k$ belongs to.


\para
It is possible to shrink $\bbeta$ towards the corresponding estimates of the external study rather than to zero, i.e. targeting shrinkage \cite[]{Gruber1998}. However, unless the two experiments are expected to be very similar in terms of design, quality, effect size distribution, and the exact meaning of the two corresponding $\beta_k$'s, this may do more harm than good. For example, our illustration on the joint use of the two methylation studies does clearly not satisfy these conditions: due to the dilution, the $\beta_k$'s in Study 2 are bound to be weaker than those in Study 1, and likely in a non-uniform way. Yet, in very well-controlled settings targeted shrinkage may be a useful extension.

\para
We end with some practical remarks. The adaptive group-regularized logistic and linear ridge procedures are implemented in the R-package \texttt{GRridge}, available via \url{www.few.vu.nl/~mavdwiel/grridge.html}. It depends on the package \texttt{penalized} \cite[]{Goeman2010b}, which is used for model fitting and cross-validation. \texttt{GRridge} provides all functionality described in this paper, including: both versions (the non-iterative, systems-based one and the iterative one), adaptive regularization on multiple partitions, variable selection, estimation of predictive accuracy by cross-validation and convenience functions to create partitions of the variables using co-data. In addition, it allows for including non-penalized variables, e.g. clinical information. It also includes both data sets discussed here. The iterative version is the default, but based on the simulations we believe one can safely use the faster, non-iterative version for $p \leq 1000$.
The iterative algorithm, however, is also fairly fast. For the first example ($p=40,000$, 7 iterations on two partitions), constructing the group-regularized ridge classifiers took 3m01s and 3m27s, for tuning the global penalty $\lambda$ by LOOCV and group-regularization, respectively. Hence, 6m28s in total on a 3GHz laptop with 3.5Mb RAM. The second example ($p=9,491$, 3 iterations on four partitions) took 31s, 23s and 14s for $\lambda$-cross-validation, group-regularization and selection, respectively, so 1m08s in total. The code used to produce the results of \texttt{GRridge} in this paper is included in the Supplementary Material.
\section{Acknowledgements}
We are particularly grateful to Tristan Mary-Huard for his critical comments and suggestions on earlier versions of this manuscript.
In addition, we thank Sanja Farkas for providing the raw data of her study \cite[]{Farkas2013} and Carel Peeters for discussing several aspects of ridge regression. This study was partly supported by the OraMod project, which received funding from the European Community under the Seventh Framework Programme, grant no. 611425. DNA methylation data of the self-samples were obtained as part of a project supported by the European Research Council (ERC advanced 2012-AdG, proposal 322986; Mass-care), by which also Wina Verlaat and Saskia M. Wilting were supported.

\bibliographystyle{author_short3} 
\bibliography{C://Synchr//Bibfiles//bibarrays}      

\pagebreak
\section*{Figures}

\end{document}